\documentclass{article}
\usepackage[utf8]{inputenc}
\usepackage{amssymb}
\usepackage{graphicx,float}
\usepackage{paralist}
\usepackage{hyperref}
\usepackage{caption,subcaption}
\usepackage{siunitx,tabularx}

\usepackage{dcolumn,booktabs,paralist,float}
\usepackage{appendix}

\usepackage{natbib}
\usepackage{amsthm}
\usepackage{mathtools} 
\usepackage[customcolors]{hf-tikz} 
\usetikzlibrary{patterns}
\usetikzlibrary{matrix,decorations.pathreplacing}
\usepackage{csvsimple} 
\usepackage{multirow}
\usepackage{thmtools, thm-restate}

\usepackage{graphicx}
\usepackage{amsmath}
\usepackage{hyperref}

\usepackage{centernot}
\usepackage{caption}
\usepackage{subcaption}
\usepackage[english]{babel}
\usepackage{algorithm}
\usepackage{algpseudocode}
\usepackage{mathtools}

\DeclarePairedDelimiterX{\expectarg}[1]{[}{]}{%
  \ifnum\currentgrouptype=16 \else\begingroup\fi
  \activatebar#1
  \ifnum\currentgrouptype=16 \else\endgroup\fi
}

\newcommand{\innermid}{\nonscript\;\delimsize\vert\nonscript\;}
\newcommand{\activatebar}{%
  \begingroup\lccode`\~=`\|
  \lowercase{\endgroup\let~}\innermid 
  \mathcode`|=\string"8000
}

\newcommand{\Ex}{\mathbb{E}}
\newcommand{\Prob}{\mathbb{P}}

\newcommand{\cov}{\mathrm{Cov}}

\usepackage{relsize}
\makeatletter
\newcommand{\distas}[1]{\mathbin{\overset{#1}{\kern\z@\sim}}}%
\newsavebox{\mybox}\newsavebox{\mysim}
\newcommand{\distras}[1]{%
  \savebox{\mybox}{\hbox{\kern1pt$\scriptstyle#1$\kern1pt}}%
  \savebox{\mysim}{\hbox{$\sim$}}%
  \mathbin{\overset{#1}{\kern\z@\resizebox{\wd\mybox}{\ht\mysim}{$\sim$}}}%
}

\pgfkeys{tikz/mymatrixenv/.style={decoration={brace},every left delimiter/.style={xshift=8pt},every right delimiter/.style={xshift=-8pt}}}
\pgfkeys{tikz/mymatrix/.style={matrix of math nodes,nodes in empty cells,left delimiter={[},right delimiter={]},inner sep=1pt,outer sep=1.5pt,column sep=2pt,row sep=2pt,nodes={minimum width=20pt,minimum height=10pt,anchor=center,inner sep=0pt,outer sep=0pt}}}
\pgfkeys{tikz/mymatrixbrace/.style={decorate,thick}}

\newcommand{\tran}{\mathsf{T}}

\makeatother

\usepackage{booktabs}

\usepackage{microtype}
\title{Scalable recommender system based on factor analysis}
\author{Disha Ghandwani\\Stanford University
\and Trevor Hastie\\Stanford University}
\date{\today}
\begin{document}
\maketitle
\begin{abstract}
Recommender systems have become crucial in the modern digital landscape, where personalized content, products, and services are essential for enhancing user experience. This paper explores statistical models for recommender systems, focusing on crossed random effects models and factor analysis. We extend the crossed random effects model to include random slopes, enabling the capture of varying covariate effects among users and items. Additionally, we investigate the use of factor analysis in recommender systems, particularly for settings with incomplete data. The paper also discusses scalable solutions using the Expectation Maximization (EM) and variational EM algorithms for parameter estimation, highlighting the application of these models to predict user-item interactions effectively.
\end{abstract}
\section{Introduction}
Recommender systems are crucial in today's tech-driven world, helping users find the right products, content, or services quickly. As online information grows rapidly, these systems need to be efficient and scalable. This paper explores how statistical models, like crossed random effects and factor analysis, can be used to build better recommender systems. We focus on models that handle large datasets with many missing entries, offering methods to improve predictions and user satisfaction. 
One of the statistical models employed in recommender systems is the crossed random effects model. Previous studies, such as those by \cite{ghos:hast:owen:2021}, \cite{ghos:hast:owen:logistic2022}, and \cite{ghandwani2023scalable}, have focused on scalable fitting of these models. Typically, in settings where $R$ users rate $C$ items, only a small fraction of the potential ratings, $N \ll R \times C$, are observed. The basic crossed random effects model captures user-specific and item-specific behavior through random effects, and it has the advantage of interpretability, allowing for inference on parameters that can be used for client-specific and item-specific recommendations.

\cite{ghandwani2023scalable} expanded on this by incorporating random slopes into the model, accounting for variability in the effect size of covariates across different users and items. However, in scenarios where covariate information is sparse or unavailable, factor analysis provides a robust alternative by uncovering hidden factors that can explain user-item interactions.

This paper explores the application of factor analysis in recommender systems, particularly for matrices with incomplete data, and discusses scalable approaches for parameter estimation using the EM algorithm. By addressing these challenges, we aim to enhance the predictive accuracy and scalability of recommender systems in real-world applications.\\

\section{Related works}
Matrix factorization has been a key technique in the development of recommender systems. \cite{5197422} introduced matrix factorization methods that significantly improved recommendation accuracy by effectively capturing latent factors in user-item interactions. \cite{NIPS2007_d7322ed7} proposed Probabilistic Matrix Factorization (PMF), which introduced a probabilistic framework to matrix factorization, offering better performance, especially in scenarios with sparse data. Further advancements were made by \cite{JMLR:v11:mazumder10a}, who developed spectral regularization algorithms for learning from large, incomplete matrices. Their work focused on enhancing the scalability and efficiency of matrix factorization methods, making them more practical for large-scale applications. \cite{candes2008exactmatrixcompletionconvex} introduced a groundbreaking approach to matrix completion by showing that it is possible to recover a low-rank matrix from a small number of observed entries using convex optimization techniques. 

These foundational studies have informed the development of more sophisticated and scalable approaches to matrix factorization, which continue to play a crucial role in modern recommender systems.

\subsection{Crossed random effects model}
A number of statistical models can be used as recommender systems and the crossed random effects model is one of them. Previous works by \cite{ghos:hast:owen:2021} and \cite{ghandwani2023scalable} focused on providing scalable fit to such models. We work in the setting where $R$ users rate $C$ items, and it is usually the case that only $N \ll R \times C$ of the ratings are observed. A simplified version of a crossed random effects model would be
\begin{equation}
y_{ij} = \beta_0+ a_i +b_j + \varepsilon_{ij}, \hspace{0.2cm} \forall i \in \{1,\cdots, R\} \hbox{   and   } j\in \{1,\cdots, C\}.
\end{equation}
Here, the random effects are $a_i \in \mathbb{R}$, $b_j \in \mathbb{R}$ and the stochastic error is $ \varepsilon_{ij} \in \mathbb{R}$. The intercept $\beta_0 \in \mathbb{R}$ and they assumed that $a_i \sim N(0, \sigma^2_a)$ , $b_j \sim N(0, \sigma^2_b), $
and $\varepsilon_{ij} \distas{i.i.d} \mathbin{N}(0,\sigma^2_e)$ are all independent. In certain cases, we may also have information on various characteristics of each (item, client) pair and the user's rating of the item. For instance, we may have information such as the user's age, gender, genre, language, and duration for online content. \cite{ghos:hast:owen:2021} considered the crossed random-effects model with random intercept terms for users and items, given by
\begin{equation}
\label{eq:intercept_model}
y_{ij} = \beta_0+ a_i +b_j+x_{ij}^{\tran}\beta + \varepsilon_{ij}, \hspace{0.5cm} \forall i \in \{1,\cdots, R\} \hbox{   and   } j\in \{1,\cdots, C\}.
\end{equation}

The crossed random effects model helps us in capturing the user-specific and item-specific behavior through the random effects; in this case, the random intercepts.
The crossed random effects model comes with an additional advantage of being interpretable and we could perform inference on $\beta$ and use it for client and item specific recommendation. \\

\cite{ghandwani2023scalable} provided scalable solution to crossed random effects model with random intercepts as well as random slopes. The model considered by them is given by 
\begin{equation}
\label{eq:random_slopes_uncompact}
y_{ij} = \beta_0+ a_i +b_j + x_{ij}^{\tran}(\beta+\tilde{a}_i+\tilde{b}_j) + \varepsilon_{ij} \hspace{0.3 cm} \forall i \in \{1,\cdots, R\} \hbox{   and   } j\in \{1,\cdots,C\}, \end{equation}
where random effects $\tilde{a}_i \in \mathbb{R}^{p}$ and $\tilde{b}_j \in \mathbb{R}^{p}$ and an error $ \varepsilon_{ij} \in \mathbb{R}$. They assumed that $\tilde{a}_i \sim N_p(0, \tilde{\Sigma}_a)$ and $\tilde{b}_j \sim N_p(0, \tilde{\Sigma}_b)$, with other terms having the same meaning. The crossed random effects model with random slopes helps us in
capturing the variability of the effect size of covariates among different users
and items.\\

It may not always be the case that we have enough information on users and items through covariates. In such cases, factor analysis models help us capture hidden factors that could make up for unknown covariates. We shall be exploring the application of factor analysis as a recommender system and start by providing a summary of factor analysis. \\
\section{Factor analysis}


Factor analysis is a powerful statistical technique used to simplify complex data sets by identifying underlying factors that explain the relationships among observed variables. By reducing a large number of interrelated variables into a smaller set of unobserved variables, or factors, this method helps reveal the hidden structure within the data. It is particularly useful for understanding the common patterns or themes that link various data points, making it easier to analyze and interpret the information. The primary aim of factor analysis is to uncover these underlying factors that contribute to the observed interdependencies among the variables, providing valuable insights into the data's underlying structure.\\

\subsection{Complete set of observations per individual}
The model aims to explain a set of $C$ observations in each of $R$ individuals using a vector of $K$ common factors ($f_{i}$), where there are fewer factors per unit than observations per unit ($K < C$). Each individual has $K$ of their common factors, which are related to the observations through the factor loading matrix ($L \in \mathbb{R}^{C\times K}$). For a single individual, the model can be represented as:
\begin{equation}
\label{eq:complete_factor_model}
\mathbf{y}_i = \mathbf{L} \mathbf{f}_i + \boldsymbol{\varepsilon}_i \end{equation}
Here, $\mathbf{y}_i $ is the set of observations for the $i$th individual, $\mathbf{f}_i $ is the vector of factors for the $i$th individual with $\cov(f_i) = I_k$, and $\boldsymbol{\varepsilon}_i$ represents the error factor with $\cov(\boldsymbol{\varepsilon}_i)  = \Psi$. Therefore, $\mathbf{y}_i$ follows a Gaussian distribution with mean zero and covariance matrix $\Psi + L L^T$ and marginal distribution of $(\mathbf{y}_i)_{i = 1}^R$ is given by 
\begin{equation}
\prod_{i=1}^R \frac{1}{\sqrt{|\Psi + L L^T|}}   \exp \left( -\frac{1}{2} {\mathbf{y}_i}^\tran \left(\Psi + L L^T \right)^{-1} \mathbf{y}_i\right) .
\end{equation}

There are a number of ways to fit obtain maximum likelihood estimates of $\Psi$ and $L$ in the factor model, principal component method and expectation maximization (EM) algorithm are a few of them. We discuss the EM algorithm below

\subsubsection{Expectation Maximization approach}
The complete likelihood for the model in $(\ref{eq:complete_factor_model})$ is given by
\begin{align*}
\prod_{i=1}^R \pi(\mathbf{y}_i |\mathbf{f}_i)&  \prod_{i=1}^R \pi(\mathbf{f}_i )\\
& \propto \prod_{i=1}^R \frac{1}{\sqrt{|\Psi|}}   \exp \left( -\frac{1}{2} (\mathbf{y}_i - \mathbf{L} \mathbf{f}_i)^\tran \Psi^{-1} (\mathbf{y}_i - \mathbf{L} \mathbf{f}_i) \right) e^{-\mathbf{f}_i^\tran \mathbf{f}_i / 2 } .
\end{align*}
where $\mathbf{f}_i$  is treated as missing observations. E-step involves computing $\Ex[\mathbf{f}_i| \mathbf{y}_i ]$ and $\cov(\mathbf{f}_i| \mathbf{y}_i )$ and the M-step involves maximizing the expected complete likelihood given $\{\mathbf{y}_i \}$ with respect to $\mathbf{L} $ and $\Psi$. The conditional distribution of $\mathbf{f}_i$ given $\mathbf{y}_i$ is given by
\begin{equation}
\mathbf{f}_i | \mathbf{y}_i \sim \mathcal{N}_K \left(\left(L^T \Psi^{-1}L + I_K \right)^{-1} L^T \Psi^{-1}\mathbf{y}_i, \left(L^T \Psi^{-1}L + I_K \right)^{-1}\right)
\end{equation}


\subsection{Haphazardly sparse observations per subject}
Factor analysis could be used as an effective tool for recommender systems. Though most of the tools are built for  complete matrices, we are here to explore the idea of factor analysis for incomplete matrices. The suggested approach could be used as an effective recommender system. The model we use is,
\begin{equation}
\label{eq:model_without_intercepts}
y_{ij} = f_i^\tran l_j + \varepsilon_{ij}, \hspace{0.5cm} \forall i \in \{1,\cdots, R\} \hbox{   and   } j\in \{1,\cdots, C\},
\end{equation}
where $ f_i, l_j \in \mathbb{R}^K$, $ f_i \sim N_K(0, I_K)$ and $\varepsilon_{ij} \overset{i.i.d}{\sim} N(0, \Psi_{jj})$ and $z_{ij} = 1$ if $y_{ij}$ is observed and zero else. The primary focus of the model is to predict the rating that future users would provide for observed items. Therefore, it makes sense to consider fixed features for items and randomness in client features.
If we have information on features of the item-client pair, given by $x_{ij}$, we can use them into the model by considering
\begin{equation}
y_{ij} =\mu_0 + x_{ij}^\tran \beta + f_i^\tran l_j + \varepsilon_{ij}, \hspace{0.5cm} \forall i \in \{1,\cdots, R\} \hbox{   and   } j\in \{1,\cdots, C\},
\end{equation}
where $x_{ij}, \beta \in \mathbb{R}^p$. We can add a column of ones into $X$ to accommodate intercept into $\beta$, i.e., the model is
\begin{equation}
\label{eq:model}
y_{ij} =x_{ij}^\tran \beta + f_i^\tran l_j + \varepsilon_{ij}, \hspace{0.5cm} \forall i \in \{1,\cdots, R\} \hbox{   and   } j\in \{1,\cdots, C\}.
\end{equation}

The goal is to provide maximum likelihood estimates of $\beta, \{l_j\}_{j=1}^C$ and $\Psi$. We attempt to provide a scalable solution for the estimation task. As already mentioned $N \ll R \times C$, therefore, we want the suggested algorithm to be scalable with the number of ratings observed. \\

Let $\mathbf{y}_i$ represent the vector of ratings provided by client $i$ (length of $\mathbf{y}_i$ is less than or equal to $R$) and $L_{i.}$ represent the factor loadings concerning the items rated by client $i$. Converting ($\ref{eq:model}$) in the matrix notation, we have
\begin{equation}
\label{eq:model_without_intercepts_matrix}
\mathbf{y}_{i} = X_{i.} \beta + L_{i.} \mathbf{f}_i + \boldsymbol{\varepsilon}_{i} .
\end{equation}
Under model ($\ref{eq:model_without_intercepts_matrix}$), $\Ex \left[\mathbf{y}_i \right] = X_{i.} \beta $ and $\cov(\mathbf{y}_i) = L_{i.} L^{\tran}_{i.} +\Psi_{i.}$ where $X_{i.}$ denotes the matrix of covariates for client $i$ and $\Psi_{i.} = \cov(\boldsymbol{\varepsilon}_{i})$.
Then, the marginal distribution of $\mathbf{y}_i$ is given by\\
\begin{equation}
f(\mathbf{y}_i) \propto \frac{1}{\sqrt{|L_{i.} L^{\tran}_{i.} +\Psi_{i.}|}} \exp{\left(-\frac{1}{2} (\mathbf{y}_i - X_{i.} \beta )^\tran (L_{i.} L^{\tran}_{i.} +\Psi_{i.})^{-1} (\mathbf{y}_i - X_{i.} \beta ) \right)}
\end{equation}
If $\mathbf{Y}$ denotes the vector of concatenated ratings of all users, then the marginal distribution of $\mathbf{Y}$ is given by the product of the marginal likelihood of $\{ f(\mathbf{y}_{i})\}_{i=1}^R$. There is no closed-form solution of the maximum likelihood estimates of $(\ref{eq:model})$. Although, there are multiple iterative algorithms to maximize the likelihood, Expectation Maximization (EM) algorithm is one of them.
EM algorithm treats $\{f_i\}_{i=1}^R$ as unobserved (missing) data and the complete likelihood represents the product of the probability density function of unobserved data and probability density function of conditional distribution of observed data given unobserved data. The complete likelihood under the model ($\ref{eq:model}$) is given by
\begin{align}
\label{eq:likelihood}
L & := \pi(\mathbf{Y} | \{f_i\}_{i=1}^R) \prod_{i=1}^R \pi(f_i)\\
& = \exp \left( -\frac{1}{2} \sum_{i=1}^R \sum_{j=1}^C z_{ij} \left[ \log(\Psi_{jj}) + \frac{ (y_{ij} - x_{ij}^\tran \beta -\mathbf{f}_i  ^\tran \mathbf{l}_j )^2}{\Psi_{jj}}\right] \right) \prod_{i=1}^R e^{- \mathbf{f}_i^\tran \mathbf{f}_i / 2 } .
\end{align}

Maximizing the complete likelihood above could be seen as minimizing a loss, which is very similar to matrix factorization with an additional penalty on $\{f_i\}_{i=1}^R$. For matrix factorization, most of the algorithms alternate between updating model parameters and providing fit for missing entries. On the other hand, we use the algorithm to estimate the model parameters based on only the observed data. The model parameters could be then used to predict future ratings. For a client-item pairing, we predict the rating by $ x_{ij}^\tran \beta + f_i ^\tran l_j$ if few ratings for both client and item have been observed. If the information has been missing for either client or item, we just use $ x_{ij}^\tran \beta$ as a prediction.\\

The other models that we considered were factor analysis with random intercepts for users and one with random intercepts for users as well as items. The model can be represented by the equations,  
\begin{equation}
\label{eq:model_intercept_client}
y_{ij} = x_{ij}^\tran \beta + f_i^\tran l_j + a_i + \epsilon_{ij}, \hspace{0.5cm} \forall i \in \{1,\cdots, R\}\hbox{   and   } j\in \{1,\cdots, C\}
\end{equation}
and 
\begin{equation}
\label{eq:model_intercepts}
y_{ij} = x_{ij}^\tran \beta + f_i^\tran l_j + a_i + b_j + \epsilon_{ij}, \hspace{0.5cm} \forall i \in \{1,\cdots, R\}\hbox{   and   } j\in \{1,\cdots, C\}, 
\end{equation}

where $a_i \in \mathbb{R}$, $ a_i \overset{i.i.d}{\sim} N(0, \sigma^2_a)$ and $b_j \in \mathbb{R}$, $ b_j \overset{i.i.d}{\sim} N(0, \sigma^2_b)$. The model ($\ref{eq:model_intercept_client}$) helps in capturing intercept effects due to users which provide assistance to the factor model (\ref{eq:model}). In the scenarios with a large number of items, we could also consider a random intercept term for items presented in the model ($\ref{eq:model_intercepts}$). For the models (\ref{eq:model_without_intercepts}), (\ref{eq:model}), and (\ref{eq:model_intercept_client}), we could use the traditional EM algorithm to estimates the parameters. For the model (\ref{eq:model_intercepts}), we use variational EM to estimate the respective parameters. The variational EM has an advantage of being scalable and it is seen to be empirically consistent.
\section{EM algorithm}{\label{sec.EM}}
Various algorithms can be used to fit factor models, and the Expectation Maximization (EM) algorithm is one of them. \cite{Ghahramani1996TheEA} illustrated the use of the EM algorithm to fit the mixture of factor models. In most of the scenarios, factor analysis has been applied to complete matrices. As the target here is to use factor model as a recommender system, we discuss the use of the EM algorithm to fit the factor model on incomplete data matrices. In the following subsection, we illustrate the application of the EM algorithm in the estimation of parameters for the model (\ref{eq:model}). For the EM algorithm, $\{f_i\}_{i=1}^{R}$ is treated as missing data. We can also simplify the algorithm under the assumption $\Psi_{jj} = \sigma^2_e, \, \, \forall j \in \{1, \cdots, C \}$ if we can be assume that the unobserved noise have an equal variance for all items.
\subsection{E-step}
In the E-step, we calculate the expected log-likelihood given the observed data, i.e., $Q : = \Ex[L|Y]$. To compute $Q$, we need to compute the conditional distribution of missing data given the observed data, i.e., $\pi (f_1, \cdots, f_R|Y)$.
\begin{align*}
\pi & (f_1, \cdots, f_R|Y) \propto \Prob(Y, \{f_i\}_{i=1}^R) \\
& \propto \exp \left \{ -\frac{1}{2} \sum_{i=1}^R \sum_{j=1}^C \frac{z_{ij} (y_{ij} - x_{ij}^\tran \beta - f_i ^\tran l_j)^2}{\Psi_{jj}} -\frac{1}{2} \sum_{i=1}^R f_i ^\tran f_i \right \}\\
& \propto \exp \Bigg \{ -\frac{1}{2} \sum_{i=1}^R f_i^\tran \left( I_k + \sum_{j=1}^{C}\frac{z_{ij}l_{j} l_{j}^{\tran}}{\Psi_{jj}} \right) f_i\\
& \hspace{2cm} + f_i ^\tran \sum_{j=1}^{C}\frac{z_{ij}(y_{ij} - x_{ij}^\tran \beta )l_{j}}{\Psi_{jj}} \Bigg\} . \\
\end{align*}
Thus, $f_i$ follow a normal distribution conditional on $Y$ with
\begin{equation}
\label{eq:mean_u_i}
\Ex[f_i|y] = \left( I_k + \sum_{j=1}^{C}\frac{z_{ij}l_{j} l_{j}^{\tran}}{\Psi_{jj}} \right)^{-1} \left(\sum_{j=1}^{C}\frac{z_{ij}(y_{ij} - x_{ij}^\tran \beta)l_{j}}{\Psi_{jj}}\right)
\end{equation}
and
\begin{equation}
\label{eq:var_u_i}
Var[f_i|Y] = \left( I_k + \sum_{j=1}^{C}\frac{z_{ij}l_{j} l_{j}^{\tran}}{\Psi_{jj}} \right)^{-1} .
\end{equation}
Expectation of the complete likelihood is given by 
\begin{align*}
\Ex[L|Y] & = -\frac{1}{2} \sum_{i=1}^R \sum_{j=1}^C z_{ij} \Ex\left[ \log(\Psi_{jj}) + \frac{ (y_{ij} - x_{ij}^\tran \beta - f_i ^\tran l_j)^2}{\Psi_{jj}}\bigg|Y\right]\\
 &\hspace{3 cm} -\frac{1}{2} \sum_{i=1}^R \Ex \left[f_i^\tran f_i\right|Y] + \text{const.}
\end{align*}
\subsection{M-step}
In the M-step, we can compute fresh estimates of $\beta, \{v_j\}_{j=1}^C$ and $\Psi$ by solving the following equations.
\begin{equation}
\frac{\partial}{\partial \beta} \Ex[L|Y] = 0,\,
\frac{\partial}{\partial l_j} \Ex[L|Y] = 0,  \text{ and } \,
\frac{\partial}{\partial \Phi_{jj}} \Ex[L|Y] = 0 \, \forall j \in \{1, \cdots, C\}.
\end{equation}
The above equations simplify to
\begin{equation}
\label{eq:beta}
\beta = \bigg(\sum_{j=1}^C \sum_{i=1}^{R}z_{ij} x_{ij}x^{\tran}_{ij}\bigg)^{-1}\left(\sum_{j=1}^C \sum_{i=1}^{R} z_{ij} \Ex(y_{ij} - f_i^ \tran l_j | Y)x_{ij}\right), 
\end{equation}
\begin{equation}
\label{eq:vj}
l_j= \bigg(\sum_{i=1}^{R}z_{ij} \Ex[f_i f_i ^\tran|Y]\bigg)^{-1}\left(\sum_{i=1}^{R} z_{ij} (y_{ij} -x_{ij}^\tran \beta )\Ex[f_i|Y]\right), 
\end{equation}
and
\begin{align}
\Psi_{jj} & = \sum_{i=1}^R \frac{z_{ij}\Ex \left[(y_{ij} - x_{ij}^\tran \beta - f_i^\tran l_j)^2 |Y\right]}{\sum_{i=1}^R z_{ij}}\\
\label{eq:Psi_j}
& = \sum_{i=1}^R \frac{z_{ij}(y_{ij} - x_{ij}^\tran \beta)^2 - z_{ij} l_j^\tran \Ex[f_i f_i ^\tran|Y] l_j }{\sum_{i=1}^R z_{ij}}.
\end{align}
As the M-step involves solving multiple linear equations, finding the exact solution could be computationally expensive. To make sure that each step is linear in $N$, we apply backfitting in each step.
\begin{algorithm}
\caption{EM algorithm for factor analysis model}\label{alg:cap}
\begin{algorithmic}
\Require $Y = \{y_{ij}\}$, $X = \{x_{ij}\}$ and dimension of factor features $k$.
\State Initialize $\beta \sim N_p(0, \mathcal{I})$ and $v_j \sim N_k(0, \mathcal{I}), \, \, \forall j \in \{1, \cdots, C \}$
\State Initialise $\Psi_{jj} \sim \chi^2_1 , \, \, \forall \, j \in \{1, \cdots, C \}.$
\While{$\beta$, $\{l_j\}_{j=1}^{C}$ and $\{\Psi_{jj}\}_{j=1}^{C}$ do not converge}
\State Compute $\Ex[f_i| Y]$ and $Var[f_i|Y]$ by (\ref{eq:mean_u_i}) and (\ref{eq:var_u_i}).
\State Update $\beta$ based on current estimates of $\{l_j\}_{j=1}^{C}$ and $\{E[f_i|Y]\}_{i=1}^R$ using (\ref{eq:beta}).
\State Update $\{l_j\}_{j=1}^{C}$ based on current estimates of $\beta$, $\{E[f_i|Y]\}_{i=1}^R$ and $\{E[f_i f_i ^T|Y]\}_{i=1}^R$ using (\ref{eq:vj}).
\State Update $\{\Psi_{jj}\}_{j=1}^{C}$ based on current estimates of $\beta$, $\{l_j\}_{j=1}^C$ and $\{E[f_i f_i ^T|Y]\}_{i=1}^R$ using (\ref{eq:Psi_j}).
\EndWhile
\end{algorithmic}
\end{algorithm}
\section{Variational EM algorithm}
Variational EM was introduced by Beal and Ghahramani (2003) and it is applicable when missing data has a graphical model structure as in the models $(\ref{eq:model_intercept_client})$ and $(\ref{eq:model_intercepts})$, i.e, $f_i$, $a_i$ and $b_j$ are correlated conditional on $Y$. 
As model $(\ref{eq:model_intercepts})$ is generalisation of model $(\ref{eq:model_intercept_client})$, we shall illustrate the algorithm to fit model $(\ref{eq:model_intercepts})$. As we are dealing with a more complicated model, we assume that $\Psi_{jj} = \sigma^2_e \, \, \forall j $, although everything will go ahead similarly without the assumption. The complete log-likelihood under the model $(\ref{eq:model_intercepts})$ is given by :-
\begin{align*}
L(\beta, \{l_j\}, f_i, a_i, b_j, & \sigma^2_a, \sigma^2_b , \sigma_e^2)  := \log \Prob(Y,f_i,a_i,b_j|\beta, \{l_j\}, \sigma^2_a, \sigma^2_b , \sigma_e^2) \\
& = -\frac{1}{2} \sum_{i=1}^R \sum_{j=1}^C z_{ij} \left[ \log(\sigma^2_e) + \frac{ (y_{ij} - x_{ij}^\tran \beta - f_i ^\tran l_j - a_i - b_j )^2}{\sigma^2_e}\right] \\
& \hspace{2cm}-\frac{1}{2} \sum_{i=1}^R f_i^\tran f_i-\frac{1}{2 \sigma_a^2} \sum_{i=1}^R a_i^2 -\frac{1}{2\sigma_b^2} \sum_{j=1}^C b_j^2
+ \text{const.}
\end{align*}

We could rewrite the complete likelihood above by using the change of variables defining $\tilde{a}_i : = a_i / \sigma_a$, i.e., $\tilde{a}_i \sim \mathcal{N}(0,1)$ and $\tilde{f}_i : = (f_i, \tilde{a}_i)$ such that $\tilde{f}_i \sim \mathcal{N}_{K + 1}(0,I_{K + 1}) $ and $\tilde{l}_j : = (l_j, \sigma_a)$. 
\begin{align*}
\label{eq:likelihood_under_intercepts}
L(\beta, \{\tilde{l}_j\}, \tilde{f}_i, b_j,  \sigma^2_b , \sigma^2_e) & := \log \Prob(Y,\tilde{f}_i,b_j|\beta, \{\tilde{l}_j\}) \\
& = -\frac{1}{2} \sum_{i=1}^R \sum_{j=1}^C z_{ij} \left[ \log(\sigma^2_e) + \frac{ (y_{ij} - x_{ij}^\tran \beta - \tilde{f}_i ^\tran \tilde{l}_j  - b_j )^2}{\sigma^2_e}\right] \\
& \hspace{2cm}-\frac{1}{2} \sum_{i=1}^R \tilde{f}_i ^\tran \tilde{f}_i  -\frac{1}{2\sigma_b^2} \sum_{j=1}^C b_j^2
+ \text{const.}
\end{align*}

and, therefore, the marginal log-likelihood will be given by
\[l(\beta, \{\tilde{l}_j\},  \sigma^2_b , \sigma_e^2) := \int L(\beta, \{\tilde{l}_j\}, \tilde{f}_i, b_j,  \sigma^2_b , \sigma^2_e) \partial \tilde{f}_i  \partial b_j. \]

Under factored variational EM, instead of finding the expectation of complete likelihood given $Y$, we substitute the conditional distribution of $(\tilde{f}_i,b_j)$ given $Y$ with $q(\tilde{f}_i ,b_j)$ belonging in a restricted set $Q$ of probability distributions.\\

We define
\begin{equation} F(q,\beta, \{\tilde{l}_j\}, \sigma^2_b , \sigma_e^2) : = \int q(\tilde{f}_i ,b_j)\log \frac{\Prob(Y,\tilde{f}_i ,b_j|\beta, \{\tilde{l}_j\},  \sigma^2_b , \sigma_e^2)}{q(\tilde{f}_i,b_j)} \partial \tilde{f}_i \partial b_j \end{equation}
which we view as an approximation to $l(\beta, \{\tilde{l}_j\},  \sigma^2_b , \sigma_e^2)$.\\

Here is the relation between $F(q,\beta, \{\tilde{l}_j\}, \sigma^2_b , \sigma_e^2)$ and $l(\beta, \{\tilde{l}_j\},  \sigma^2_b , \sigma_e^2)$,
\begin{align*}
F(q,\beta, \{\tilde{l}_j\}, \sigma^2_b , \sigma_e^2) = l(\beta, \{\tilde{l}_j\},  \sigma^2_b , \sigma_e^2) - \mathrm{KL} \left(q|| \Prob(\tilde{f}_i ,b_j|Y, \beta, \{\tilde{l}_j\},  \sigma^2_b , \sigma_e^2)\right).
\end{align*}
For the model (\ref{eq:model_intercepts}),we choose
\begin{align*}Q & = \bigg\{ q: q(\tilde{f}_1,\cdots, \tilde{f}_R, b_1,\cdots , b_C) \\
& \hspace{4cm}= q_f(\tilde{f}_1,\cdots, \tilde{f}_R)  q_b(b_1, \cdots, b_C)\bigg\}\end{align*}
If $Q = \{q: q(\alpha) = \prod q_u(\alpha_u)\}$, it follows from Beal (2003) that for the exponential family, optimal update for distribution of each component of $\alpha$ is given by
\begin{equation}
q_u^{(k)}(\alpha_u) \propto \Prob \left(Y, \alpha_u| \alpha_{-u} = \Ex_{q^{(k-1)}} (\alpha_{-u})\right)
\end{equation}
Variational EM possesses the property that
\begin{equation}F(q^{(k-1)},\theta^{(k-1)}) \leq F(q^{(k)},\theta^{(k-1)}) \leq F(q^{(k)},\theta^{(k)}) \end{equation}
\subsection{Variational E-step}
In the Variational E-step, we calculate $  \Ex_q[  L(\beta, \{\tilde{l}_j\}, \tilde{f}_i, b_j,  \sigma^2_b , \sigma^2_e)]$ which requires computing the optimal distributions for $\{\tilde{f}_i\}_{i=1}^R$ and $\{b_j\}_{i=1}^R$. The respective distributions in the variational E-step are given by
\[ q^{(k)}_f(\tilde{f}_1, \cdots, \tilde{f}_R) = \prod_{i=1}^R q^{(k)}(\tilde{f}_i) \quad \mbox{ and } \quad q^{(k)}_b(b_1, \cdots, b_C) = \prod_{j=1}^C q^{(k)}(b_j) \]
with
\begin{equation*}q^{(k)}(\tilde{f}_i) = N(\mu^{(k)}_{f,i}, \Sigma^{(k)}_{f,i}) \quad \mbox{and} \quad q^{(k)}(b_j) = N(\mu^{(k)}_{b, j},{\sigma^2}^{(k)}_{b, j}), \end{equation*}
where
\begin{equation}
\label{eq:var_mu_b}
\mu^{(k)}_{b,j} ={\sigma^2}^{(k)}_{b, j} \left(\frac{\sum_{i=1}^{R}z_{ij}\big(y_{ij} - x_{ij}^{\tran}\beta^{(k-1)}  - {\mu^{(k-1)}_{f,i}}^\tran \tilde{l}_j^{(k-1)}\big)}{\left(\sigma^{(k-1)}_e\right)^{2}}\right)
\end{equation}
and 
\begin{equation}
\label{eq:var_sigma_b}
{\sigma^2}^{(k)}_{b, j} = \left(\frac{\sum_{i=1}^R z_{ij}}{\left(\sigma^{(k-1)}_e\right)^{2}} + \frac{1}{\left(\sigma^{(k-1)}_b\right)^{2}}\right)^{-1}.
\end{equation}
Also, 
\begin{equation}
\label{eq:var_mu_f}
\mu^{(k)}_{f,i} = \Sigma^{(k)}_{f,i} \left(\frac{\sum_{j=1}^{C}z_{ij}\big(y_{ij} - x_{ij}^{\tran}\beta^{(k-1)}  - \mu^{(k-1)}_{b,j} \big){\tilde{l}}^{(k-1)}_j}{\left(\sigma^{(k-1)}_e\right)^{2}}\right)
\end{equation}
and
\begin{equation}
\label{eq:var_sigma_f}
\Sigma^{(k)}_{f,i} = \left(\frac{\sum_{j=1}^C z_{ij} {\tilde{l}}^{(k-1)}_j {\tilde{l}}^{(k-1)}_j}{\left(\sigma^{(k-1)}_e\right)^{2}} + I_{K + 1} \right)^{-1}.
\end{equation}
\subsection{M-step}
In the M-step, we maximize the expected log likelihood computed in variational E-step to  compute the fresh estimates of $\beta, \{\tilde{l}_j\}_{j=1}^C$, $\sigma^2_b$ and $\sigma^2_e$. As $\tilde{l}_j = (l_j, \sigma_a)$, we maximize the expected log likelihood with respect to $\beta, \{l_j\}_{j=1}^C$, $\sigma^2_a$, $\sigma^2_b$ and $\sigma^2_e$. As the expected log likelihood is a convex, maximizing the function is equivalent to equating the derivative with respect to each parameter to zero. 
\begin{equation*}
\frac{\partial}{\partial \beta} F(q,\beta, \{\tilde{l}_j\}, \sigma^2_b , \sigma_e^2) = 0,\, \, \,
\frac{\partial}{\partial l_j} F(q,\beta, \{\tilde{l}_j\}, \sigma^2_b , \sigma_e^2) = 0
\end{equation*}
\[
\frac{\partial}{\partial \sigma_a} F(q,\beta, \{\tilde{l}_j\}, \sigma^2_b , \sigma_e^2) = 0\]
\begin{equation*}
\frac{\partial}{\partial \sigma^2_b} F(q,\beta, \{\tilde{l}_j\}, \sigma^2_b , \sigma_e^2) = 0,\, \, \,
\frac{\partial}{\partial \sigma^2_e} F(q,\beta, \{\tilde{l}_j\}, \sigma^2_b , \sigma_e^2) = 0
\end{equation*}
\begin{equation}
\label{eq:var_m_beta}
\hat \beta ^{(k)}= \bigg(\sum_{j=1}^C \sum_{i=1}^{R}z_{ij} x_{ij}x^{\tran}_{ij}\bigg)^{-1}\left(\sum_{j=1}^C \sum_{i=1}^{R} z_{ij} \left(y_{ij} - {\mu^{(k)}_{f,i}}^ \tran {\tilde{l}}_j^{(k-1)} - \mu^{(k)}_{a,i} - \mu^{(k)}_{b,j}\right)x_{ij}\right)
\end{equation}
\begin{equation}
\label{eq:var_m_vj}
\hat l_j^{(k)}= \bigg(\sum_{i=1}^{R}z_{ij} \Ex_q[u_i u_i ^\tran]\bigg)^{-1}\left(\sum_{i=1}^{R} z_{ij} \left(y_{ij} -x_{ij}^\tran \hat \beta ^{(k)} - \mu^{(k)}_{a,i} - \mu^{(k)}_{b,j} \right)\Ex_q \left[u_i\right]\right)
\end{equation}
\begin{equation}
\label{eq:var_m_sigma2e} \sigma^2_e = \frac{1}{N} \sum_{i=1}^R \sum_{j=1}^C z_{ij} \Ex_q \left[ (y_{ij} - x_{ij}^\tran \beta - u_i^\tran v_j - a_i - b_j )^2 \right]\end{equation}
\begin{equation}
\label{eq:var_m_sigma2a} \sigma^2_a = \frac{1}{R} \sum_{i=1}^R \Ex_q [a_i a_i^\tran] = \frac{1}{R} \sum_{i=1}^R \left(\mu^{(k)}_{a,i}{ \mu^{(k)}_{a,i}}^\tran + \Sigma^{(k)}_{a,i} \right)\end{equation}
\begin{equation}
\label{eq:var_m_sigma2b}
\sigma^2_b = \frac{1}{C} \sum_{j=1}^C \Ex_q [b_j b_j^\tran] = \frac{1}{C} \sum_{j=1}^C \left(\mu^{(k)}_{b,j} {\mu^{(k)}_{b,j}}^\tran + \Sigma^{(k)}_{b,j} \right)\end{equation}
Similar to before, M-step involves solving multiple linear equations and therefore, finding the exact solution could be computationally expensive. To make sure that each step is linear in $N$, we apply backfitting in each step.
\begin{algorithm}
\caption{Factored Variational EM algorithm for factor analysis model}\label{alg:cap}
\begin{algorithmic}
\Require $Y = \{y_{ij}\}$, $X = \{x_{ij}\}$ and dimension of factor features $k$.
\State Initialize $\beta \sim N_p(0, \mathcal{I})$ and $v_j \sim N_k(0, \mathcal{I}), \, \, \forall j \in \{1, \cdots, C \}$
\State Initialise $\sigma^2_a, \sigma^2_b, \sigma^2_e \sim \chi^2_1.$
\While{$\beta$, $\{v_j\}_{j=1}^{C}$, $\sigma^2_a$, $\sigma^2_b$, and $\sigma^2_e$ do not converge}
\State Update the optimal distributions by (\ref{eq:var_mu_b}), (\ref{eq:var_sigma_b}), (\ref{eq:var_mu_f}) and (\ref{eq:var_sigma_f}).
\State Update $\beta$, $\{v_j\}_{j=1}^{C}$, $\sigma^2_a$, $\sigma^2_b$, $\sigma^2_e$ by (\ref{eq:var_m_beta}), (\ref{eq:var_m_vj}), (\ref{eq:var_m_sigma2e}) ,
(\ref{eq:var_m_sigma2a}), (\ref{eq:var_m_sigma2b}).
\EndWhile
\end{algorithmic}
\end{algorithm}
\section {Recommendation on Movie Lens}
We performed the real data analysis using ``Movie lens 100K dataset" available on $\hyperlink{https://www.kaggle.com/datasets/prajitdatta/movielens-100k-dataset}{kaggle}$.The datasets describe ratings and free-text tagging activities from MovieLens, a movie recommendation service. It contains around 100,000 ratings and from 942 users on 1682 movies. Each user has rated at least 20 movies. The dataset also comes with certain information on users and movies like age, gender, and occupation of users and genre of the movies. We use the combined information of user and movies as features for linear model, i.e., $x_{ij} = (\text{age}_i, \text{gender}_i, \text{one hot vector for genre}_j)$. The dataset comes with five splits of the dataset into training and test datasets. We use the same splits to compare our algorithm on factor analysis with and without random intercepts to the softImpute algorithm developed by \cite{JMLR:v16:hastie15a} and the baseline intercept model~(\ref{eq:intercept_model}). We use mean square error on predicted ratings as our metric for comparison. The softImpute algorithm does not use the information on covariates, so we perform a linear regression of ratings on covariates and apply the softImpute algorithm on residuals to make the comparisons valid.
The experiments can be replicated by using the provided seed values. We fit the following models on the training data:-
\begin{enumerate}
\item \textbf{Model with just client intercepts: } \label{Model_1}
\[y_{ij} = x_{ij}^\tran \beta + a_i + \epsilon_{ij}, \hspace{0.5cm} \forall i \in \{1,\cdots, R\}, \hspace{0.3cm} j\in \{1,\cdots, C\}. \]
\item \textbf{Model with just client and item intercepts: } \label{Model_2}
\[y_{ij} = x_{ij}^\tran \beta + a_i + b_j + \epsilon_{ij}, \hspace{0.5cm} \forall i \in \{1,\cdots, R\}, \hspace{0.3cm} j\in \{1,\cdots, C\}. \]
\item \textbf{Factor model without intercepts: } \label{Model_3}
\[y_{ij} = x_{ij}^\tran \beta + f_i^\tran l_j + \epsilon_{ij}, \hspace{0.5cm} \forall i \in \{1,\cdots, R\}, \hspace{0.3cm} j\in \{1,\cdots, C\}. \]
\item \textbf{Factor model with client intercepts: } \label{Model_4}
\[y_{ij} = x_{ij}^\tran \beta + f_i^\tran l_j + a_i + \epsilon_{ij}, \hspace{0.5cm} \forall i \in \{1,\cdots, R\}, \hspace{0.3cm} j\in \{1,\cdots, C\}. \]
\item \textbf{Factor model with client and item intercepts: } \label{Model_5}
\[y_{ij} = x_{ij}^\tran \beta + f_i^\tran l_j + a_i + b_j + \epsilon_{ij}, \hspace{0.5cm} \forall i \in \{1,\cdots, R\}, \hspace{0.3cm} j\in \{1,\cdots, C\}. \]
\end{enumerate}
\begin{table}
\centering
\begin{tabular}{|l|l|l|l|l|l|l|}
\hline
& Hard Impute & Model \ref{Model_1} & Model \ref{Model_2} & Model \ref{Model_3} & Model \ref{Model_4} & Model \ref{Model_5}\\
First split & 0.8953 & 	1.0930 & 	0.9068 & 	0.9477 & 	{\color{red} 0.8716 }& 	0.9133 \\
Second split & {\color{red} 0.8823} & 	1.0640 & 	0.8893 & 	0.9246 & 	0.8938 & 	0.8906 \\
Third split & 0.8776 & 	1.0346 & 	0.8763 & 	0.8967 & 	{\color{red} 0.8492 } & 	0.8590 \\
Fourth split & 0.8705 & 	1.0411 & 	0.8751 & 	0.8659 & {\color{red} 0.8403 } & 	0.8580 \\
Fifth split & 0.8640 & 	1.0438 & 0.8749 & 	0.8806 & 	{\color{red} 0.8639 } & 	0.8729 \\
\hline
\end{tabular}
\caption{Results on Movie lens dataset}
\label{Tab:stitch_fix}
\end{table}
For the softImpute package, we tried hard impute ($\lambda = 0$) for different number of latent features, $K$ varying from 1 to 3. Similary for models \ref{Model_3}, \ref{Model_4} , and \ref{Model_5}, we explore different number of latent features ranging from 1 to 3. For each model, we report the results with minimum mean square error. For model \ref{Model_3}, minimum mean square error is obtained using $K = 3$ latent features and for model \ref{Model_4}, minimum mean square error is obtained using $K = 2$ latent features for most of the data splits. For model \ref{Model_5}, the result is consistent with model \ref{Model_4}.
\bibliographystyle{apalike}
\bibliography{bigdata}

\begin{thebibliography}{}

\bibitem[Candes and Recht, 2008]{candes2008exactmatrixcompletionconvex}
Candes, E.~J. and Recht, B. (2008).
\newblock Exact matrix completion via convex optimization.

\bibitem[Ghahramani and Hinton, 1996]{Ghahramani1996TheEA}
Ghahramani, Z. and Hinton, G.~E. (1996).
\newblock The em algorithm for mixtures of factor analyzers.

\bibitem[Ghandwani et~al., 2023]{ghandwani2023scalable}
Ghandwani, D., Ghosh, S., Hastie, T., and Owen, A.~B. (2023).
\newblock Scalable solution to crossed random effects model with random slopes.

\bibitem[Ghosh et~al., 2022a]{ghos:hast:owen:2021}
Ghosh, S., Hastie, T., and Owen, A.~B. (2022a).
\newblock Backfitting for large scale crossed random effects regressions.
\newblock {\em Annals of Statistics}, 50(1):560--583.

\bibitem[Ghosh et~al., 2022b]{ghos:hast:owen:logistic2022}
Ghosh, S., Hastie, T., and Owen, A.~B. (2022b).
\newblock Scalable logistic regression with crossed random effects.
\newblock {\em Electronic Journal of Statistics}, 16(2):4604--4635.

\bibitem[Hastie et~al., 2015]{JMLR:v16:hastie15a}
Hastie, T., Mazumder, R., Lee, J., and Zadeh, R. (2015).
\newblock Matrix completion and low-rank svd via fast alternating least
  squares.
\newblock {\em Journal of Machine Learning Research}, 16(104):3367--3402.

\bibitem[Koren et~al., 2009]{5197422}
Koren, Y., Bell, R., and Volinsky, C. (2009).
\newblock Matrix factorization techniques for recommender systems.
\newblock {\em Computer}, 42(8):30--37.

\bibitem[Mazumder et~al., 2010]{JMLR:v11:mazumder10a}
Mazumder, R., Hastie, T., and Tibshirani, R. (2010).
\newblock Spectral regularization algorithms for learning large incomplete
  matrices.
\newblock {\em Journal of Machine Learning Research}, 11(80):2287--2322.

\bibitem[Mnih and Salakhutdinov, 2007]{NIPS2007_d7322ed7}
Mnih, A. and Salakhutdinov, R.~R. (2007).
\newblock Probabilistic matrix factorization.
\newblock In Platt, J., Koller, D., Singer, Y., and Roweis, S., editors, {\em
  Advances in Neural Information Processing Systems}, volume~20. Curran
  Associates, Inc.

\end{thebibliography}
\end{document}